\documentclass[sigconf, nonacm]{acmart}

\AtBeginDocument{%
  }

\begin{document}

\title{EducaSim: Interactive Simulacra for CS1 Instructional Practice}

\author{Cameron Mohne}
\email{mohnec1@cs.stanford.edu}
\authornote{Both authors contributed equally to this research.}
\affiliation{%
  \institution{Stanford University}
  \state{California}
  \country{USA}
}
\orcid{0000-0002-6399-1329}

\author{Nicholas Vo}
\authornotemark[1]
\email{nickvo@cs.stanford.edu}
\affiliation{%
  \institution{Stanford University}
  \state{California}
  \country{USA}
}

\author{Dora Demszky}
\email{ddemszky@stanford.edu}
\affiliation{%
  \institution{Stanford University}
  \state{California}
  \country{USA}
  }

\author{Chris Piech}
\email{piech@cs.stanford.edu}
\affiliation{%
  \institution{Stanford University}
  \state{California}
  \country{USA}
  }

\renewcommand{\shortauthors}{Mohne et al.}

\begin{abstract}
   Role play is a high-impact mode of training that has demonstrated its effectiveness in improving learning outcomes. However, it is difficult to scale to teacher instruction due to its inherent dependency on providing personnel who are both trained and available to facilitate this learning environment. This poses a challenge, especially to massive online courses which may employ and aid hundreds to thousands of novice teachers. While simulating students isn't a new approach, past work struggles with simulations being in-context, having consistent agent behavior, and having relevant modes of interaction. In this work, we present \textit{EducaSim}: a novel framework that uses generative agents to simulate a small-group section for teachers-in-training to practice instruction. \textit{EducaSim} works by implementing diverse pedagogical-based personas, actual course material, and agent-based architectures constructed for instructional practice to provide a pedagogically rich environment for teachers-in-training to engage in role play learning -- without the costly overhead that comes with it. We share our experiences with constructing and making the tool available for experimental training and preparation in a six-week CS1 course supporting 20,000 students with the study being centered on 150 out of the 1,300 volunteer teachers. We found that teachers who engaged generally saw it as a positive experience. We believe that \textit{EducaSim} is an important step to providing experiential teaching practice at scale for closely-defined settings and has great potential for future applications.
\end{abstract}



\begin{teaserfigure}
    \centering
    \includegraphics[width=0.7\textwidth]{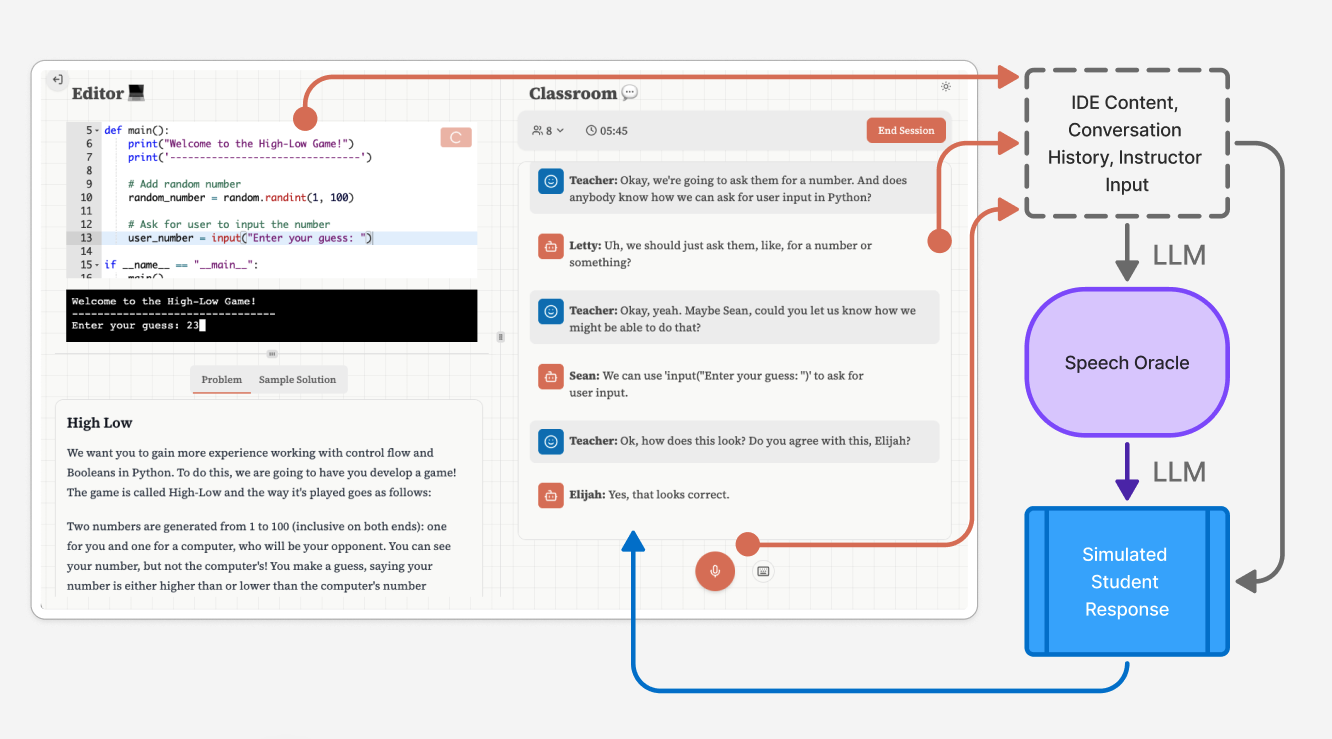}
    \caption{\textit{EducaSim} simulation environment. The user can engage through voice or text. After input, the simulation context and conversation is used to decide the next speaker and generate a student response.}
    \Description{A screen showing an example of a user interacting in the simulated environment, with a diagram of how the simulation flow works.}
    \label{fig:teaser}
\end{teaserfigure}

\maketitle

\section{Introduction}
Teacher training is a vital, yet difficult process that is currently stricken with issues in scale, cost, and implementation \cite{hill2024effectiveness}. Role play remains a well-recognized teaching and coaching technique to actively engage learners. Beyond the ability to enable coaching and reflection, it can be especially helpful for preparing teachers for the myriad of classroom scenarios they may encounter and allow them to practice verbalizing their lesson \cite{shaikh2024rehearsal, Lavanya2024roleplay, moreno2020roleplay, yue2025mathvcllmsimulatedmulticharactervirtual}.

However, scaling role play can be challenging in the context of large courses that focus on human-centered learning such as Code in Place \cite{codeinplace} and Schoolhouse \cite{schoolhouse}, which operate at a scale of up to thousands of teachers. There exists an enormous time and cost constraint that cannot be met due to the limited number of trained teachers available to facilitate this learning at various times across the world. Advances in Large Language Models (LLMs) have allowed agents utilizing generative text to better simulate human dialogue and actions without extrinsically specified goals \cite{park2023generativeagentsinteractivesimulacra}. Synthesizing this research poses a novel opportunity to democratize teacher coaching through role play with simulated agents and time for feedback and self-reflection. Importantly, these simulations offer a 24/7, low-cost alternative to live teacher coaching.

To investigate the interdisciplinary potential between advances in agents and education and juxtapose them with limitations of prior work, we built and released \textit{EducaSim} as an optional preparation device to teachers for an introductory computer science course. We provide three exercises rooted in common classroom scenarios and actual course content and assess the general uptake and use of the tool.

In this paper, we present our methodology, architecture, results, and an in-depth analysis on how instructors used and viewed their experiences with the tool. Our work highlights the potential of simulated student agents as a scalable training and coaching tool for pre-service teachers.

\subsection{Related Work}
Past works have utilized LLM-based agents as tools for teacher coaching whether it be for classroom management, teaching pedagogy, or other aspects \cite{markel2023gpteach, wang2024tutorcopilothumanaiapproach, shaikh2024rehearsal, yue2025mathvcllmsimulatedmulticharactervirtual}. These LLM-based agents continue to improve and are used in a multitude of use-cases \cite{guo2024largelanguagemodelbased, wang2024survey} where they are specifically constructed and interact in their respective environments \cite{park2023generativeagentsinteractivesimulacra, wei2025improvingparallelprogramperformance,delriochanona2025generativeaiagentsbehave}. Despite this, simulations intended for coaching still fall short in areas such as understanding specific domains \cite{ZHENG2025100990, wang2024tutorcopilothumanaiapproach}, agent behavior \cite{markel2023gpteach, ZHENG2025100990}, and generalization \cite{shaikh2024rehearsal}.

There is a significant amount of work loosely in this domain primarily due to simulations being an immersive format that incites the fundamental pedagogues of \textit{role play} \cite{moreno2020roleplay, Lavanya2024roleplay}. Role play is generally used in education as it helps keep students engaged \cite{Acharya2018, brown2022, crow2015}, give contextual practice \cite{cumbria1409}, tackle situations from multiple perspectives \cite{westrup2013}, and develop soft skills \cite{brown2022}. Usage spans fields from aviation to medical sciences since high-fidelity simulations allow for practice in safe, yet realistic environments \cite{ZHENG2025100990, crow2015, brown2022}---enabling experiential learning \cite{Butvilofsky2012, Rayner03072014, THEELEN201914}. This form of learning is thought to be a viable way to prepare learners for future challenges \cite{ferguson2017using, kaufman2016enhancing, mcgarr2021use}.

In the past, we've seen several approaches to scaling role play using LLMs. However, many approaches are limited and challenges still exist to creating a robust, reliable simulation. For example, GPTeach is a tool for an introductory computer science course that allows for teachers to engage in simulated group sessions with students \cite{markel2023gpteach}. However, this tool only allows for text interaction and ignores important context such as the virtual classroom environment and past student knowledge. Furthermore, there is no element of feedback incorporated to the tool. Existing platforms such as Character.ai or widely-used foundation models do not allow for a unified solution that combines classroom context, defined personas, and decision-making frameworks. As a result, the agents can often hallucinate or exhibit sycophantic behavior \cite{xu2024largelanguagemodelseducation, ZHENG2025100990, simSchool}.

Research has shown that role play and feedback are powerful tools for improving both teaching and learning outcomes. The Rehearsal system, for example, employs LLMs to simulate conflict scenarios, enabling users to practice conflict resolution in a controlled setting \cite{shaikh2024rehearsal}. This approach allows learners to experiment with different strategies and receive AI-generated feedback during the session, helping them refine their skills for real-world applications. As applied in medical training, we have seen that scenario-based roleplay, followed by feedback using Pendleton's Model, significantly improves medical students' communication and teamwork skills \cite{Lavanya2024roleplay}. Students found this method more effective than lectures, with structured feedback playing a key role in boosting competence and confidence \cite{crow2015}.

In our research, we aim to explore the general perception, use cases, and potential of contextually-grounded generative agents within educational simulacra. We present our findings in this paper.

\subsection{Main Contributions}
\textbf{(1) A novel student-agent architecture} synthesizing:
\begin{itemize}
    \item Generative agents with their own personas and memories which impact when and how they engage.
    \item Domain-contextualized agent memories to align interactions with actual course material.
    \item A classroom dynamic tree utilizing LLM as a judge.
\end{itemize}

\textbf{(2) Extended modes of user interaction in the simulation environment} in addition to standard text input, which are:
\begin{itemize}
    \item A runnable Python IDE which has code that is passed into the agent memory stream
    \item Low-latency speech-to-text transcription
\end{itemize}

\textbf{(3) Outcomes from large-scale access}.
\begin{itemize}
    \item Usage from deployment in an online CS1 course with 1,300 volunteer teachers of varying backgrounds.
    \item Sentiments from teachers on usefulness and believability
\end{itemize}

\section{Methodology}
In this segment we will go over the various components of the agents, the setup of the simulation, post-session feedback, and the associated costs for running our simulations.

\subsection{Agent Architecture}
\subsubsection{Student Personas}
We deviate from defining personas with sensitive demographic identity traits (i.e. gender, race, age) \cite{park2023generativeagentsinteractivesimulacra} to be mindful of the bias represented in the LLM. \cite{tan2025unmaskingimplicitbiasevaluating, rossi2024problems, venkit2025taleidentitiesethicalaudit, bu2025investigationvaluemisalignmentllmgenerated}.

Instead, we considered the types of behavior that are most important to represent within the training simulation. These behaviors include when and how a student engages \cite{trowler2010student, wong2022student}, the type and style of utterances (namely, how much understanding they expose and how), and the level of understanding of course material. Given this definition, we include within the student persona the following traits: (1) name, (2) engagement style, and (3) speech style. Understanding of course material is covered in more depth in the memory segment of the paper below.

\subsubsection{Memory}
\begin{figure}[h!]
    \centering
    \includegraphics[width=\linewidth]{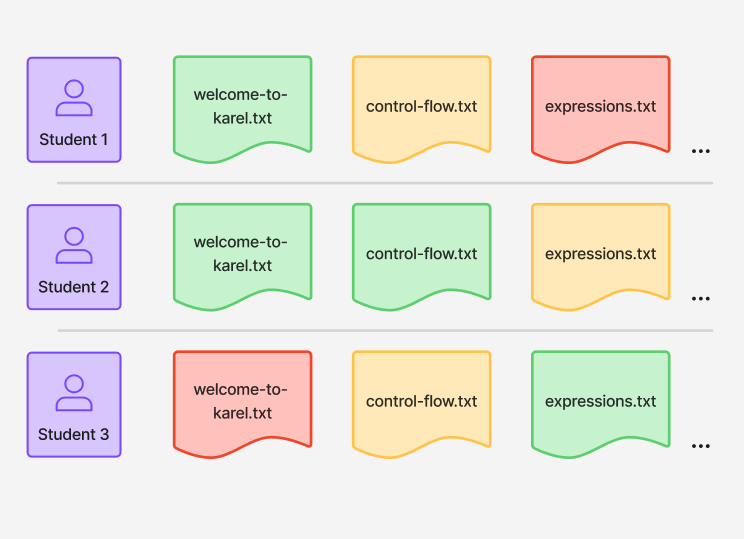}
    \caption{Student agent memories are populated with course content at varying levels of engagement.}
    \label{fig:memories}
\end{figure}

Agents are constructed with an individualized node-based memory system \cite{park2023generativeagentsinteractivesimulacra} with an initial set of foundational memories based on their personas. These randomized background memories are added to allow for realistic responses to questions beyond the course material. Then, we take text-based materials such as lecture transcripts from a course---the materials should be ordered chronologically based on when a student would encounter it, so teachers can refer to "past" material---and chunk them. Before assigning memories based on these course contents, we attach a level of engagement---low, medium, or high---which describes the integrity of interaction with each individual document. We prepend a string based on the engagement level into the then formulated memories specifically describing what the student does and does not know. This step is what allows us to define different "knowledge" states within our student agents. The end-state of this process when we run a simulation can be visualized in Figure \ref{fig:memories}.

When memories need to be retrieved---such as when they are speaking---we use a retrieval method that metricizes memories using (1) time of creation, (2) importance, and (3) semantic similarity to an anchor input text. Importance is static and generated by an LLM upon creation. To gauge similarity, we generate embeddings using an embedding model and calculate the cosine similarity to the vector embeddings of the memory. Using this scoring function, we retrieve the top-$k$ most relevant memories \cite{park2023generativeagentsinteractivesimulacra}.

\subsubsection{Decision-Making Framework}
To generate the simulated student responses, we follow a framework that helps constrain the student response. The response can be of two types: (1) an error if the agent does not have the knowledge required to respond or (2) a success if the agent has the knowledge required to respond. This framework builds on past research such as Bridge \cite{wang-etal-2024-bridging} to determine a set of archetypes for student responses based on their cognitive process. We contribute a set of responses classified by error types and success types as outlined in Table \ref{tab:types}.

\begin{table}[h!]
    \centering
    \begin{tabular}{l|l}
        \textbf{Error}   & \textbf{Success} \\ \hline
        guess & short \\
        misinterpret & well-formed \\
        careless & vague \\
        right-idea & question \\
        imprecise & \\
        question & 
    \end{tabular}
    \caption{List of error and success types.}
    \label{tab:types}
\end{table}

\begin{figure}[h!]
    \centering
    \includegraphics[width=0.75\linewidth]{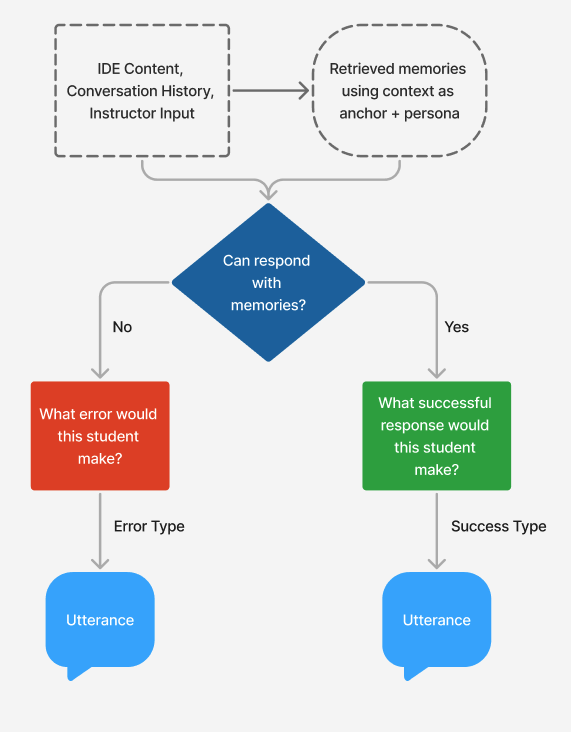}
    \caption{Student agents undergo a decision process using relevant context when tasked to generate a response.}
    \label{fig:decisiontree}
\end{figure}

To determine if an agent will either make an error or succeed, we first prompt an LLM with the environment context, retrieved memories, and persona to decide if they can reasonably reply to the last message. Based on this decision, we prompt an LLM with the same context to generate the specific success or error type; notably, the prompt for success and error are separate. Finally, using this generated response type, we prompt an LLM using the original context to generate the final utterance that the agent will create. Figure \ref{fig:decisiontree} shows a visual representation of this pipeline.

\subsection{Simulation Setup}
\subsubsection{Modes of Interaction}
Given the context of Computer Science education, we implemented a runnable IDE \cite{runno_2025} and voice-to-text via Whisper \cite{openai_whisper_v20240930}. Both modes have their outputs converted into text; the former is piped into the memory stream and the latter is treated the same as normal text conversation. While we allow for both voice and text user input, the default-facing form of communication is voice-to-text and we advise users to utilize speech when possible. This is important as we want to encourage users to practice verbalizing their instruction.

\subsubsection{Speech Oracle}
To decide which agent should speak in the simulation, we currently employ an oracle by using an LLM as a judge \cite{yue2025mathvcllmsimulatedmulticharactervirtual}. The oracle works by first combining information from the ongoing simulation, such as conversation history and instructor input, with a list of the participating student agents and their associated personas to form the context. Using this context, we then prompt an LLM from the perspective of a teacher to decide who is the most likely to respond, if anyone should respond, or if the whole class should respond. Notably, it is possible for no students to respond, in which case, the text that appears states that the entire class is silent.

\subsection{Session Feedback}
\subsubsection{Talktime Statistics}
We run a simple pass over all the messages sent in the transcript and total the number of words from each speaker. We show the total percentage of space each speaker took up---including how much silence there was. We also show a visual representing whether or not each individual agent spoke in the session.

\subsubsection{Feedback and Highlights}
Feedback generation is accomplished through prompting a separate LLM with structured instructions designed to analyze teacher-student interaction transcripts. Specifically, the LLM is prompted to systematically identify key instructional behaviors or statistics to generate the feedback. Some of the behaviors and statistics it looks at includes teacher uptake of student contributions, the proportion of teacher versus student talk, the quality and types of teacher questioning, and evidence of student misconceptions or reasoning processes, inspired by \cite{wang-etal-2024-bridging} and \cite{Demszky2023}.

Subsequently, the LLM employs cognitive task analysis techniques inspired by expert decision-making frameworks. It explicitly infers underlying reasons behind observed instructional challenges or student misunderstandings, selects targeted remediation strategies (e.g., simplifying problems, asking guiding questions, providing conceptual explanations, or prompting reflection), and clearly articulates the pedagogical intentions behind each strategy. Next, the LLM-generated feedback is structured in a constructive and actionable format. It summarizes instructional strengths, identifies specific areas for improvement supported by concrete examples from the transcript, and provides targeted suggestions for instructional enhancement. Finally, we ask the LLM to analyze the transcript for key highlights of what the teacher did well---namely things like answering questions, revoicing student understandings, or attempts at engaging students.

\subsubsection{Self-Reflection}
Reflective prompts are included to facilitate teacher self-assessment and deeper pedagogical reflection. This prompting-based approach enables automated yet contextually sensitive feedback aligned with expert instructional practices \cite{biencinto2021psychometric, ilie2024dynamics, Myllykoski-Laine02092024, article_teacher_supports}.

\subsection{Associated Costs}
\textit{EducaSim} utilizes GPT 4.1-mini and Whisper-1 for text generation and audio transcription respectively. The cost-per-session currently can range from \$0.05 - \$0.10 depending on duration. Hosting the site and all of the LLM API calls---excluding the domain cost---totals less than \$5 for 150 users at the time of writing.

\section{Results}
\subsection{Usage}
We deployed \textit{EducaSim} on its own domain as a web app due to technical limitations constraining the timeframe of integration. Focused on material from the partnered open-access CS1 course, our platform supported three exercises focused on weeks three, four, and six of the program. Each week featured a different scenario and problem to discuss in line with the week's contents. It was offered as an informal, optional preparation tool for their 1,300 teachers who serve about 20,000 students with exercises being released about a week before the respective exercise's designated week. We provide an overview of usage following the tool's deployment in Table \ref{tab:stats_overview}.

\begin{table}[h!]
    \centering
    \begin{tabular}{l|l}
        \textbf{Metric} & \textbf{Observed Value} \\
        \hline
        \# Sessions & 254 \\ 
        Mean Session Duration & 15 minutes 45 seconds \\ 
        Median Session Duration & 11 minutes 4 seconds
    \end{tabular}
    \caption{Overview of EducaSim's usage after deployment for two weeks.}
    \label{tab:stats_overview}
\end{table}

\subsection{Feature Comparison}
We offer a comparative analysis of \textit{EducaSim} features versus current alternatives that allow for simulated learning. Specifically, we compare to Character.ai \cite{character_ai}, ChatGPT \cite{openai_chatgpt}, GPTeach \cite{markel2023gpteach}, Rehearsal \cite{shaikh2024rehearsal}, and Clinical Mind AI \cite{stanford_clinicalmindai}. While many of these platforms are not specifically created for instructional training, this comparison exists to help outline the unique features and opportunities that \textit{EducaSim} offers.

\subsubsection{Modes of Interaction}
Voice-to-text is a large contributing factor towards realism as online teachers typically engage via voice. Character.ai, ChatGPT, and Clinical Mind AI support voice as a form of input, but all other listed tools which are solely created for education purposes support only text. Given the deployment context, EducaSim has an embedded and runnable IDE which can be directly piped to the agent's memories as if they were viewing the output. We will discount the models without any focus/support on coding. After honing in on the remaining relevant tools: ChatGPT and GPTeach, ChatGPT partially allows for an IDE of sorts, but it's neither runnable nor ideal for the type of simulation conducive for learning. GPTeach has no version of an IDE.

ChatGPT has one leg up in the ability that it can technically freely process images for visual input. No other tool supports this, but pedagogical usefulness needs further study since there is no promise that such a feature is integratable in its current state. GPTeach supported an imaginary Zoom call for its agents and Character.ai has a similar visualization for their various agents, which is also a leg up on us for now. The other tools do not support this.

\subsubsection{Agent Contextualization}
From prior literature, a major pain point for simulations was agents lacking domain-specific knowledge. That is, if student agents are learning about coding, they should ask questions and have memories contextualized in that particular domain. ChatGPT has a memory component for conversations, but you would always need to do some number of calls prior to set it up. Things also get muddy for cases with multiple agents. The other tools have some level of very specific context they usually work off of. There is a general difficulty overall with prior tools in creating tools flexible for new contexts in an educational setting since the education-forward tools aren't often made with flexibility in mind whilst flexible tools aren't centered solely on educational frameworks and structure. \textit{EducaSim} has an admin-side interface where existing teaching materials can be uploaded and chunked to directly create in-context agents.

\subsubsection{Speaker Selection}
Unique to \textit{EducaSim} is the ability to interact with multiple students who each possess their own respective personas and memories. While GPTeach and prior versions of Character.ai allowed for multi-agent interaction, these agents were limited in scope to single prompts as opposed to a more complex state and memory management system. Furthermore, the next speaker was decided by random choice or static heuristics which could detract from the realism of a teaching interaction. For example, if a student was specifically called upon, the same student should be expected to answer in the next turn. Or, if no student is expected to respond, the conversation should reflect that. \textit{EducaSim} addresses these nuanced cases with an LLM-based oracle that decides who the next speaker should be. As a result, we are able to create a simulation that more reasonably mirrors a real conversation.

\subsubsection{Feedback}
Feedback is a vital part to enabling learning from simulations. Both Rehearsal and Clinical Mind AI support feedback through defined rubrics or LLM-generated responses. Rehearsal allows for in-session, single-turn feedback while Clinical Mind AI allows instructors to customize the feedback students receive at each stage of their interaction. \textit{EducaSim} supports post-session feedback and combines defined statistics (e.g. talktimes), LLM-generated feedback, and self-guided reflection questions to allow for learners to learn from their session.

\subsection{User Reactions}
The feedback from optional forms was notably positive. On the teachers-only forum in a thread centered around the tool's release, one teacher discussed their thoughts on the feedback: 
\begin{quote}
    "I tried this simulation twice.  The first time, I had no success in getting the "students" to speak.  After reviewing the feedback, I had more success the 2nd time by saying I recognized a number of old students from before but saw a number of new faces.  Then I got a few replies.  I also shared that I had sometimes hesitated to speak up last year.  I added that I soon felt more comfortable participating when I got to know the group.  More answers...      (so some good hints from feedback)"
\end{quote}
While this in particular is limited to the microcosm presented in the simulation, it highlights that coaching and a change in approach impacts how the simulated class acts. On this notion, there was considerable discussion on the classroom dynamics:
\begin{quote}
    "This is a great tool! definitely more advanced than GPTeach. That section is pretty tough. Would love to see a section that gets the learner talk time into the 'Great' region."
\end{quote}
being one from a teacher who had actually tested GPTeach \cite{markel2023gpteach}; and then a different returning teacher had stated
\begin{quote}
    "It's really a great tool for practice. However, my actual section is quite interactive this time. Last year's section was just the same as the simulation."
\end{quote}
on the topic of the first exercise in which the students are too shy to speak and often opt for silence.

\section{Discussion and Practical Uses}
Through user studies, we have found an interesting number of potential use cases, which we share here.

\subsection{Teacher Training Supplement}
As applied in our deployment of \textit{EducaSim}, we see this as a way for teachers, especially in settings like Code in Place \cite{codeinplace} or Schoolhouse \cite{schoolhouse} as a supplement to traditional training which is significantly harder to scale as the number of teachers increases. For these types of online courses, the simulated environment can be much more similar as there may be less physical cues to draw on. In fact, some real sessions from these platforms have been reported to be entirely text-based. These simulations can serve as high-frequency training that leaves more time for feedback iteration between follow-up synchronous sessions with coaches.

\subsection{Teacher Preparation}
The second use case is that these simulations can be used by teachers, especially in online settings, to prepare for upcoming lessons as a way to refresh their own understanding of concepts and engagement strategies. In end-of-week recap threads, reflections from some active users honed in on which students were engaging in the lesson, how they were engaging, and what specific strategies were or weren't working when it came time to engage students. Using these reflections, an instructor could construct a simulation to practice responding to these specific situations. Interestingly, while we cannot state causation, the sudden shift compared to prior weeks may indicate a shift in how they decided to reflect.

\subsection{Student Learning}
One exciting use case for \textit{EducaSim} is to serve as student assignments. Whether it be for assessment or to gauge student understanding, prior work shows that such instances with human-to-human interactions could be beneficial \cite{malik2024learners}. Given users have to guide agents to a solution, it may be possible to gain insight on the user's own understanding of material based on how they lead the interactions.

\subsection{Teacher Screening Component}
On a similar notion to student learning, \textit{EducaSim} could be one step in the screening process for evaluating teachers who are prepared to teach online courses. Currently, the requirements for various online teacher screenings involve a walkthrough demo, solving some problems, and/or giving debugging advice \cite{codeinplace, schoolhouse}. These completely asynchronous activities are vulnerable to Generative AI simply doing the work---outside of demos, which are inherently scripted by the teacher. Tuning agents with some level of randomness in personality could potentially provide non-uniform experiences that break out of this mold with more distinct markers (e.g. time spent per utterance) for potential dishonesty.

\section{Limitations and Future Work}
\subsection{Student Personas}
We acknowledge that such agents can not perfectly represent student behavior; rather, they are intended to be an approximation of what an instructor might encounter during instruction. Similar to how peers might role play as students in instructor training, the simulated students are intended to be a proxy for practicing instruction. Humans are incredibly multi-dimensional and attunement for perceived pedagogical usability isn't perfect. We feel it is vital to highlight how we are intentionally missing out on key identity traits such as cultural context, neurodivergence, or personal perception that can play a significant role in a classroom environment but cannot be ethically represented using a LLM. In selecting agent names in an effort to sample a global audience similar to one represented in our partner, we also acknowledge that such names may contain bias. Furthermore, our framework cannot fully represent the diversity of individual human understandings, confusions, or questions that a student may have. Thus, \textit{EducaSim} may currently be more suited towards developing novice teachers as opposed to well-experienced teachers who may need more specialized practice.

\subsection{Speaker Oracle}
Using LLM as a judge performs well in a confined space, but classrooms are complex with many modes of interaction. There exist common teaching strategies that can confuse the architecture, such as \textit{popcorning} where students, pick other students to speak, or settings like thank-pair-shares. Robustness can be improved and added support for multi-turn dialogue is an area of improvement.

\subsection{Text as Data}
Our current architecture is reliant on text. It inherently loses a lot of signal on tone and other markers for understanding such as visual cues. It also constrains what inputs and outputs are available for the agents. For a generalizable, comprehensive solution, more work needs to be done to increase the input and output modality outside the scope of text output and speech and text input.

\subsection{Future Work}
Teachers have used \textit{EducaSim} to practice rehearsing their lesson, anticipating student questions, and reviewing their content understanding. However, this learning is dependent on experience and uptake of the generated feedback. Some instructors may need a more structured learning experience through the simulation such as through in-session hints, direction, or highly specific scenarios. Additionally, isolation poses a challenge to instructors in MOOCs. Adding features for collaborative sessions and peer-generated feedback could enrich the experience through social, peer-learning.

Beyond the current platform, an exciting future direction is to make the experience more multi-modal. Two features to add to the learning experience includes (1) allowing agents to respond with audio and (2) adding visual elements that more closely replicate the session such as Zoom room windows and speaking indicators. To make the tool more generalizable, we would also like to add materials from different subjects and have distinct, creatable "courses" by coaches (or teachers in the case of usage as a tool for student learning). We believe that teachers possess a rich knowledge of learning scenarios, and a platform that allows such teachers to create these scenarios as simulations could be a powerful tool.

\section{Conclusion}
We present \textit{EducaSim} as an exciting tool that allows for experiential instructional practice to happen at scale. Leveraging multiple agents with individualized contextual memories, thought out interaction frameworks, and the generalization power of LLMs, \textit{EducaSim} is a solution that allows instructors to practice meaningful aspects of teaching such as explaining concepts, engaging students, responding to a range of questions, or noticing gaps in learning -- all at a low-cost of \$0.05 - \$0.10 per session. In the context of a massive open online course that employs thousands of volunteer instructors, we found that the sentiment surrounding the tool was positive, underlying the potential for the tool to be used beyond an optional supplement. With continual improvement, we believe that simulated role play paired with actionable feedback is a powerful learning tool that can lead to a future where instruction can drastically be improved at scale.

\begin{acks}
    This work was supported by a grant from the Stanford Accelerator for Learning. We thank Daniel Lee, Julia Markel, Allen Nie, Mei Tan, Joon Sung Park, Mike Hardy, and our testers for their feedback and discussion. We are also grateful to the Code in Place team who allowed us to interface with their teachers and team for wide-scale testing, simulation direction, and feedback.
\end{acks}

\newpage
\bibliographystyle{ACM-Reference-Format}
\bibliography{bibfile}

\end{document}